\documentclass[lettersize,journal]{IEEEtran}
\usepackage{amsmath,amsfonts}
\usepackage{algorithmic}
\usepackage{algorithm}
\usepackage{array}
\usepackage[caption=false,font=normalsize,labelfont=sf,textfont=sf]{subfig}
\usepackage{textcomp}
\usepackage{stfloats}
\usepackage{url}
\usepackage{verbatim}
\usepackage{graphicx}
\usepackage{cite}
\usepackage{textcomp}
\usepackage{bm}
\usepackage{multirow}
\begin{document}

\title{Adaptive Wavelet Filters as Practical Texture Feature Amplifiers for Parkinson’s Disease Screening in OCT }

\author{Xiaoqing~Zhang, Hanfeng~Shi, Xiangyu Li, Haili~Ye, Tao~Xu, Na~Li, Yan~Hu, Fan Lv, Jiangfan Chen and Jiang Liu, \IEEEmembership{Senior~Member,~IEEE}

\thanks{Xiaoqing~Zhang and Na~Li are Center for High Performance Computing and Shenzhen Key Laboratory of Intelligent Bioinformatics, Shenzhen Institute of Advanced Technology, Chinese Academy of Sciences, Shenzhen, China}
\thanks{Xiaoqing~Zhang, Hanfeng~Shi, Yan~Hu, and Jiang Liu are Research Institute of Trustworthy Autonomous Systems and Department of Computer Science and Engineering, Southern University of Science and Technology, Shenzhen 518055, China. (Xiaoqing~Zhang and Hanfeng~Shi contribute equally; Corresponding author: Xiaoqing~Zhang and Jiang~Liu; E-mail: xq.zhang2@siat.ac.cn, liuj@sustech.edu.cn). }
\thanks{Xiangyu Li, Tao~Xu, Fan Lv, and Jiangfan Chen are with the State Key Laboratory of Ophthalmology, Optometry and Vision Science, Wenzhou Medical University, Wenzhou, China}
\thanks{Haili~Ye is Centre for Computational Science and Mathematical Modelling, Coventry University, Coventry, UK}
\thanks{Jiang~Liu is also School of Computer Science, University of Nottingham Ningbo China, Ningbo 315100, China; School of Ophthalmology and Optometry, Wenzhou Medical University, Wenzhou 325035, China; Department of Electronic and Information Engineering, Changchun University, Changchun 130022, China.}
\thanks{This paper was produced by the IEEE Publication Technology Group. They are in Piscataway, NJ.}
\thanks{Manuscript received April 19, 2021; revised August 16, 2021.}}

\markboth{Journal of \LaTeX\ Class Files,~Vol.~14, No.~8, August~2021}%
{Shell \MakeLowercase{\textit{et al.}}: A Sample Article Using IEEEtran.cls for IEEE Journals}


\maketitle

\begin{abstract}
Parkinson's disease (PD) is a prevalent neurodegenerative disorder globally. The eye's retina is an extension of the brain and has great potential in PD screening. Recent studies have suggested that texture features extracted from retinal layers can be adopted as biomarkers for PD diagnosis under optical coherence tomography (OCT) images. Frequency domain learning techniques can enhance the feature representations of deep neural networks (DNNs) by decomposing frequency components involving rich texture features. Additionally, previous works have not exploited texture features for automated PD screening in OCT. Motivated by the above analysis, we propose a novel Adaptive Wavelet Filter (AWF) that serves as the Practical Texture Feature Amplifier to fully leverage the merits of texture features to boost the PD screening performance of DNNs with the aid of frequency domain learning. Specifically, AWF first enhances texture feature representation diversities via channel mixer, then emphasizes informative texture feature representations with the well-designed adaptive wavelet filtering token mixer. By combining the AWFs with the DNN stem, AWFNet is constructed for automated PD screening. Additionally, we introduce a novel Balanced Confidence (BC) Loss by mining the potential of sample-wise predicted probabilities of all classes and class frequency prior, to further boost the PD screening performance and trustworthiness of AWFNet. The extensive experiments manifest the superiority of our AWFNet and BC over state-of-the-art methods in terms of PD screening performance and trustworthiness.
\end{abstract}


\section{Introduction}
\label{sec:intro}
\IEEEPARstart{P}{arkinson's} disease (PD) is the second most common neurodegenerative disorder globally, involving motor and non-motor manifestations~\cite{ thakur2023automated}. PD occurs with progressive loss of dopaminergic neurons in the substantia nigra, which is attributed to the pathological accumulation of $\alpha$-synuclein within cells~\cite{lee2022multimodal, robbins2021characterization}. It is also a chronic and age-related disease, and its prevalence will increase sharply due to global aging~\cite{tysnes2017epidemiology, soni2024update}. Multiple neuroimaging techniques have been employed to detect dopaminergic degeneration to investigate PD progression, especially magnetic resonance imaging (MRI). However, MRI techniques have limitations in watching extranigral pathologies of early-stage PD due to the long course that might be insufficient for precise yet early PD screening. Additionally, MRI examination is expensive and time-consuming, causing few PD patients to take this examination. Subsequently, many patients with PD miss critical intervention and treatment stages, imposing a heavy burden on individuals and society.

\begin{figure}
     \begin{minipage}{0.40\linewidth}
        \centering
        \includegraphics[width=0.95\linewidth]{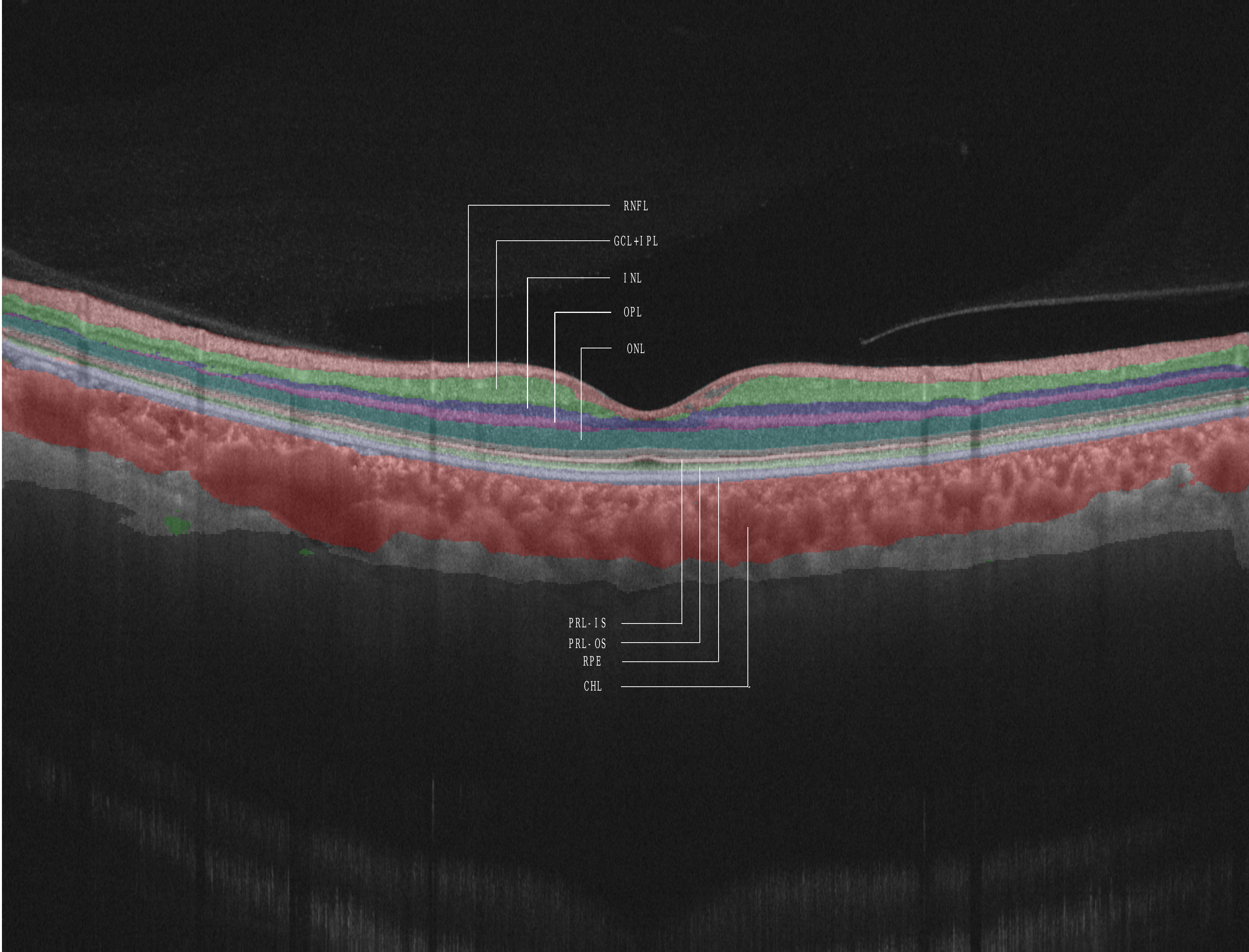} 
        (a)
    \end{minipage}
    \begin{minipage}{0.55\linewidth}
        \centering
        \includegraphics[width=0.96\linewidth]{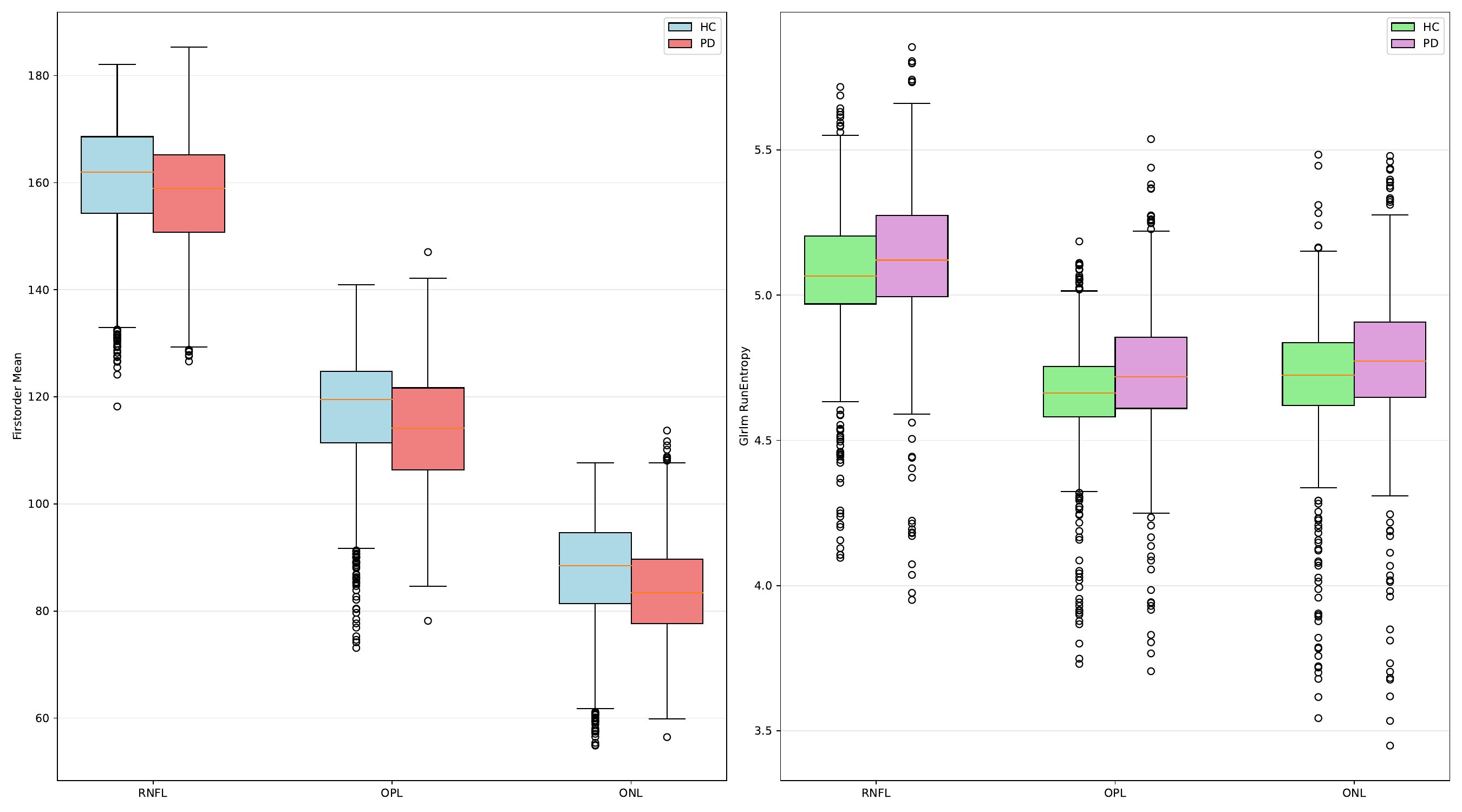}
        (b)
    \end{minipage}
    \hfill
    \begin{center}
     \begin{minipage}{0.90\linewidth}
        \centering
        \includegraphics[width=0.96\linewidth]{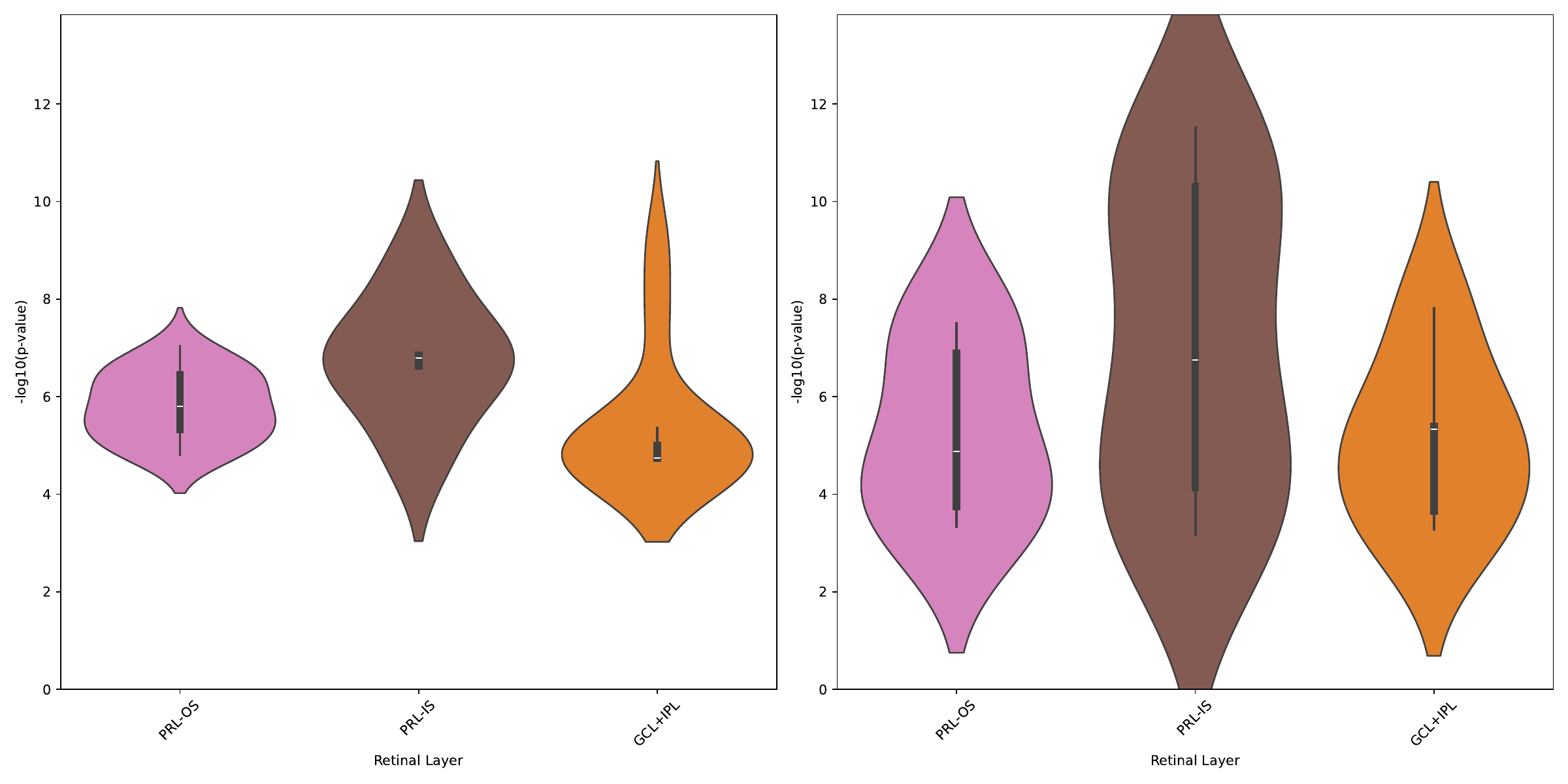}
        (c)
    \end{minipage}        
    \end{center}
    \caption{(a) An OCT image with annotated retinal nerve fiber layer (RNFL), ganglion cell layer and inner plexiform layer (GCL+IPL), inner nuclear layer (INL), outer plexiform layer (OPL), outer nuclear layer (ONL), photoreceptors inner segment (PRL-IS), photoreceptors outer segment (PRL-OS), retinal pigment epithelium (RPE), choroid layer (CHL). (b)
    The texture feature value distribution differences of first-order mean and gray level size zone matrix (GLSZM) feature-run entropy (RE) between the PD group and the healthy control group. These two texture features are extracted from RNFL, OPL, and ONL. (c) The visualizations of the correlation coefficient distributions among texture features extracted from different retinal layers based on OCT images and the frequency components extracted from high-level feature maps, which are obtained from ResNet (Left) and AWFNet (Right) accordingly. Here, the vertical axis denotes correlation coefficient distribution, and the horizontal axis denotes different retinal layers. }
    \label{fig1}
\end{figure}

The eye's retina is considered as another window to brain pathology due to the same embryo origins, offering a significant means to diagnose PD~\cite{alves2023structural, de2018eye}. Recently, optical coherence tomography (OCT) imaging techniques have gradually been utilized to investigate PD pathogenesis on the retina issues through a non-invasive yet cheap manner~\cite{huang2021central, chorostecki2015characterization}, as presented in Fig.~\ref{fig1}(a). Ortuño-Lizarán et al.~\cite{ortuno2020dopaminergic} found the degeneration of dopaminergic cells and their dendritic plexus in retinal tissue from PD patients. Unlu et al. \cite{unlu2018correlations} found the significant thinning of different retinal layer thicknesses between the PD group and healthy control (HC) group, such as retinal nerve fiber layer (RNFL), inner plexiform layer (IPL), inner nuclear layer (INL), outer plexiform layer (OPL), and outer nuclear layer (ONL). Hasanov et al.~\cite{hasanov2019functional} reported that functional and morphological differences are associated with PD. Jordan et al.~\cite{alves2023structural} shown that structural and functional changes of retinal layers can serve as potential biomarkers for early PD diagnosis through morphological features and texture features. Motivated by existing clinical works, Richardson et al.~\cite{richardson2024multimodal} applied a convolutional neural network (CNN) to recognize PD. Shi et al.~\cite{shi2024retinal} proposed a retinal structure guidance-and-adaption network (RSGANet) for automated early PD recognition. However, previous works showed that obtaining promising PD screening performance is still challenging.

According to our extensive survey, we found an intriguing phenomenon is that existing works mainly focus on extracting morphological features from the retina issue, for investigating structural changes and functional disorders of PD, \textbf{ignoring texture features as other pathological manifestation forms of PD progress in the macro perspective.} This phenomenon is associated with intrinsic etiopathogenesis, such as chemical changes, $ \alpha$-synuclein accumulation, and loss of dopaminergic neurons \textbf{from the micro view.} To investigate the meaningful differences of texture features between the PD group and the HC group visually, Fig.~\ref{fig1}(b) offers the first-order mean and gray level size zone matrix (GLSZM) feature-run entropy (RE) value distributions extracted from three different retinal layers: RNFL, OPL and ONL. We observe significant differences between the two groups, manifesting that texture features can play critical roles in PD diagnosis by serving as biomarkers.

Deep neural networks (DNNs) have achieved promising progress in automated disease recognition based on OCT images~\cite{xiao2024multi,chen2023fit,shen2023structure,yang2023aim}, such as convolutional neural networks (CNNs), vision transformer (ViT), attention-based networks, and visual foundation models. For instance, Shen et al. \cite{shen2023structure} proposed a Structure-Oriented Transformer to classify retinal diseases. Xiao et al.~\cite{xiao2024multi} proposed a multi-style spatial attention network to recognize cortical cataract severity levels. However, these OCT-oriented DNNs lack the specific ability to capture informative texture feature representations from high-level feature maps. Recently, frequency domain learning methods have gradually been applied to boost the generalization ability of DNNs by decomposing different frequency components~\cite{yang2022dual}. \textbf{Here, we suggest that these frequency components contain rich texture feature representations but have been under-explored, not to mention to boost PD screening performance in OCT.}

Inspired by the systematic analysis, we propose a novel yet effective Adaptive Wavelet Filter (AWF) as the practical texture feature amplifier, which leverages the potential of texture features in high-level feature maps of DNNs with the aid of frequency domain learning techniques. To be specific, our AWF comprises two main components: channel mixer and adaptive wavelet filtering token mixer (AWF-Mixer), as shown in Fig.~\ref{fig3}(b). The channel mixer is devised to enhance the diversities and refinement of texture feature representations. Following it, the well-designed AWF-Mixer is applied to obtain useful texture representations from high-level feature maps via triple operators: Wavelet Domain Mapping, Adaptive Group Linear Weighting, and Inverse Wavelet Transform Reconstruction. Consequently, we combine our proposed AWFs with the mainstream DNN stem to construct an Adaptive Wavelet Filtering Network (AWFNet) for automated PD screening based on OCT images, as shown in Fig.~\ref{fig3}(a). Fig.~\ref{fig1}(c) offers the visualizations of the correlation coefficients among texture features extracted from three different retinal layers based on OCT images and the frequency components extracted from high-level feature maps of DNNs: ResNet18 and our AWFNet. The visualization results show that our proposed AWFs enhance informative texture feature representations, supporting our hypothesis.

Additionally, trustworthiness is a critical yet fundamental issue in artificial intelligence (AI) based medical diagnosis applications, which might limit their real deployment~\cite{chen2024dynamic,zou2024confidence}. One straight reason is that existing DNNs have difficulties in keeping consistencies of actual predicted probabilities and confidence scores by employing the classical cross-entropy (CE) loss to optimize their learnable weights. This is mainly because CE typically treats each training sample equally in a mini-batch, unavoidably ignoring the generated confidence miscalibration errors of all classes, especially for classes with few samples. To alleviate the trustworthiness issue and further improve PD screening performance, we propose a novel Balanced Confidence (BC) Loss to leverage the potential of predicted probabilities of different classes and the corresponding training label frequency prior. The main contributions of this paper are summarized as follows:

\begin{itemize}
 \item We propose a novel AWF as the Practical Texture Feature Amplifier, highlighting helpful texture feature representations from high-level feature presentations in DNNs. Furthermore, we construct the AWFNet for automated PD screening by integrating the AWFs into the DNN stem.
\item We develop a Balanced Confidence (BC) loss to improve the trustworthiness and boost the PD screening performance of AWFNet, which attempts to fully exploit the potential of predicted probabilities of different classes and training label frequency priori. 
\item The conducted experiments verify the effectiveness and generalization ability of our AWFNet and BC through comparisons to advanced DNNs and loss methods on three OCT datasets. Further analysis from the age-related aspect manifests the superiority and trustworthiness of our proposed methods.
\end{itemize}

\section{Related Work}
\textbf{Automated OCT-based  disease recognition via DNNs.}
Over the past years, DNNs have significantly prompted the development of medical image analysis, including OCT-based disease diagnosis. Takahiro et al.~\cite{sogawa2020accuracy} ensembled nine representative CNNs, e.g., VGG16, ResNet50, and Xception, to recognize different retinal diseases such as high myopia (HM), myopic choroidal neovascularization (mCNV), and retinoschisis (RS). Shin et al.~\cite{shin2021deep} proposed a fused CNN to detect glaucoma.  Huang et al.~\cite{huang2023lesion} utilized the pretrained ResNet34 to detect myopic traction maculopathy (MTM). Zhou et al.~\cite{zhou2024gamnet} proposed a gated attention mechanism network for automated MTM grading. Chen et al.~\cite{chen2023fit} developed a feature interaction transformer to diagnose pathologic myopia automatically. In addition, automated PD recognition based on ophthalmic images has attracted the growing research interests from a broader perspective, providing a potential yet universal means for early PD diagnosis. For example, Huang et al.~\cite{HUANG2024108368} proposed a wavelet-based selection-and-recalibration network to detect PD on OCT images by emphasizing morphological features. Tran et al. \cite{tran2024deep} applied a DNN to predict PD.
Unluckily, those previous methods overlooked the important roles of texture features. In contrast, this paper focuses on enhancing texture feature representations in high-level feature maps of DNNs to improve PD screening results.

\textbf{Frequency domain learning in DNNs.}
Frequency domain analysis has been a meaningful means in the fields of signal processing and computer vision. Notably, scholars found that classical frequency domain analysis methods also can improve generalization and feature representation capabilities of DNNs~\cite{huang2023adaptive, chen2024frequency, li2024raffesdg, li2023enhancing}. Wang et al.~\cite{wang2018packing} applied the discrete cosine transformation (DCT) method to accelerate the training of DNNs.  Qin et al.~\cite{qin2021fcanet} utilized DCT to obtain multi-scale frequency components and then embedded them into the channel attention block for boosting the feature representation learning. Mishra et al.\cite{9144534} employed the wavelet transform method to decompose feature representations of DNNs into different frequency components for improving image quality. Yang et al.~\cite{9933881} incorporated different frequency components in spatial attention and channel attention block by the DWT method for improving classification performance. Unlike existing purposes of frequency domain learning utilizations in DNNs, this paper aims to apply the frequency domain learning method to serve a practical texture
feature amplifier, by decomposing high-level feature representations into different frequency components, thereby boosting PD screening results in OCT.

\textbf{Loss modification for trustworthiness and classification improvement.}
In this paper, we focus on reviewing recent loss modification methods, which are more related to our proposed method. Guo et al. \cite{guo2017calibration} added a temperature scaling parameter to CE to improve the network calibration. Liang et al. \cite{liang2020imporved} proposed the auxiliary loss named DCA by computing the difference between accuracy and predicting probability values.  Mukhoti et al.~\cite{mukhoti2020calibrating} found that FCL not only improves classification results but also benefits confidence calibration. Tao et al.~\cite{ tao2023dual} proposed a dual focal (D-Focal) loss for network calibration by considering the predicted logits corresponding to different classes and the largest predicted logits ranked after them. In contrast to prior efforts, this paper proposes the BC loss to improve the trustworthiness of DNNs by considering both predicted probabilities and training label frequency distributions of different classes.

\section{Methodology}

\subsection{General Overview of AWFNet}
\begin{figure*}
    \centering
    \includegraphics[width=0.96\linewidth]{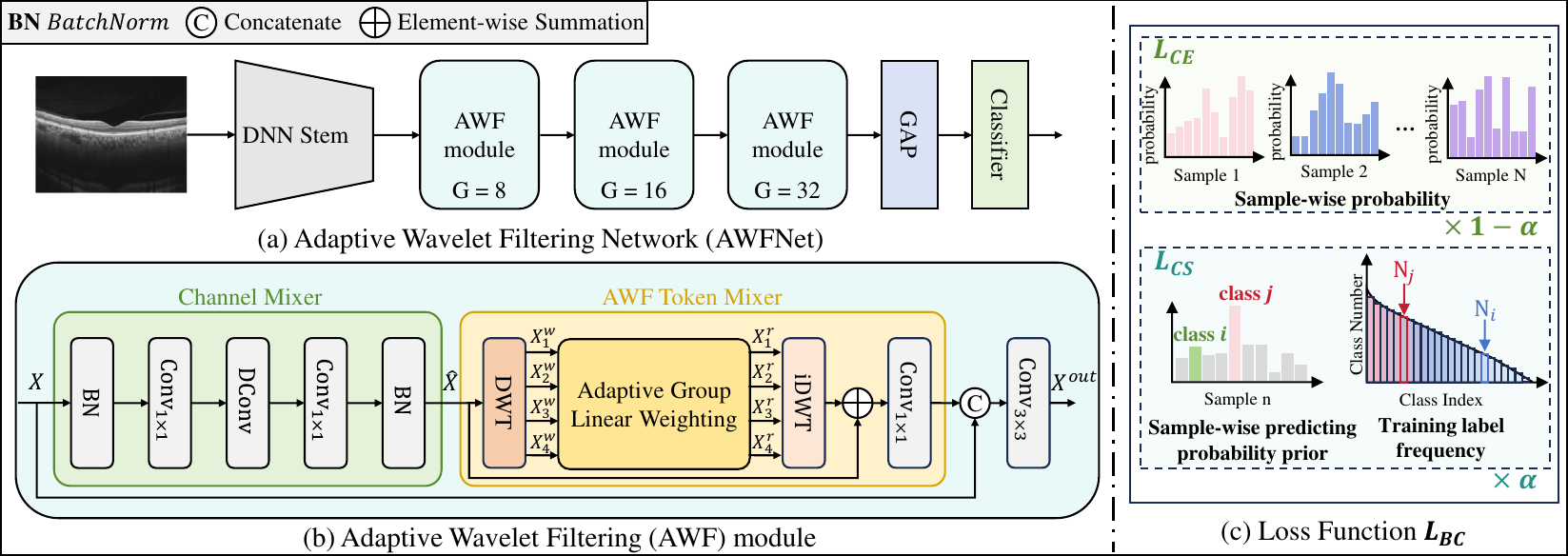}
    \caption{The general architecture of Adaptive Wavelet Filtering Network (AWFNet) with the Balanced Confidence Loss (BC) $L_{BC}$ for trustworthy PD screening under OCT images, which emphasizes important texture feature representations in high-level feature maps by the well-designed AWF. Particularly, we design a novel BC loss to boost the trustworthiness of AWFNet and further improve PD screening performance. Therefore, our AWFNet can produce trustworthy PD screening results in the inference stage, which has great potential to be deployed on real medical devices.}
    \label{fig3}
\end{figure*}

In this paper, we develop an Adaptive Wavelet Filtering Network (AWFNet) to boost PD screening performance based on OCT images by fully highlighting texture feature representations in high-level feature representations with the help of frequency domain learning. Specifically, we design an adaptive wavelet filter (AWF) serve as the texture feature amplifier, as presented in Fig.~\ref{fig3}(b), consisting of two main blocks: channel mixer and adaptive wavelet filtering token mixer (AWF-Mixer). The channel mixer is applied to enhance the diversities of feature maps for refining feature representations, and the AWF-Mixer is designed to strengthen helpful texture feature representations of high-level feature maps. Consequently, we combine AWFs with mainstream DNN stem to construct an Adaptive Wavelet Filtering Network (AWFNet) for automated PD screening based on OCT images,as offered in Fig.~\ref{fig3}(a).

Additionally, to boost the trustworthiness of our AWFNet and further improve its PD screening performance, we develop a novel balanced confidence loss to explore the potential of predicted probabilities and training label frequency distributions of different classes, aiming to optimize the learnable parameters of AWFNet effectively, as offered in Fig.~\ref{fig3}(c).

\subsection{Adaptive Wavelet Filter}
Our adaptive wavelet filtering (AWF) is supposed to enhance the extraction of informative texture feature representations in the high-level feature maps $X \in R^{B \times C \times H \times W}$ produced by the DNN stem (B is the batch size, C is the channel number, $H \times W$ is the resolution of the feature map). The AWF consists of two main blocks: channel mixer and AWF-Mixer, which will be described step by step.

\subsubsection{Channel Mixer}
To better refine feature representations of high-level feature maps through feature map diversity enhancement, we propose a channel mixer (CM), which is a combination of two point-wise convolution (Conv$1\times1$) operations, two batch normalization (BN) operations,  and a depthwise convolution (DConv) operation by following previous works~\cite{sandler2018mobilenetv2}. The detailed formulas of CM are as follows:
\begin{equation}
\hat{X} = BN(Conv_{1\times1}(DConv(Conv_{1\times1}(BN(X)))),
    \label{eq1}
\end{equation}
where $\hat{X} \in \mathbb{R}^{B \times C \times H \times W}$ denotes refined high-level feature maps, and the kernel size for DConv operation is $3\times 3$. The main reason to adopt $3\times 3$ kernel size is that it is helpful to enrich texture feature representation learning.

\subsubsection{Adaptive Wavelet Filtering Token Mixer}
Following the CM, we propose an adaptive wavelet filtering token mixer (AWF-Mixer) to enhance informative texture feature representations in the refined high-level feature maps via three independent operators: wavelet domain mapping, adaptive group linear weighting, and inverse wavelet transform reconstruction. Here, the token is another perspective to understand the refined feature maps, and each token is denoted as $x \in R^{B \times 1 \times 1 \times C}$. Token mixing aims to capture the dependencies among spatial locations in DNNs, allowing feature representation to transmit across the spatial dimensions locally or globally. In this paper, token mixing is applied to construct interactions among different frequency maps.

\paragraph{\textbf{Wavelet Domain Mapping}} 
Frequency domain learning methods have widely been utilized to improve the representational capability of DNNs. Unlike previous ones, this paper aims to apply frequency domain learning to emphasize texture feature representations in the refined feature maps to boost PD screening performance. Therefore, we propose a Wavelet Domain Mapping (WDM) operator, which decomposes the refined feature maps $\hat{X}$ into four different yet complementary frequency maps with the discrete wavelet transform (DWT) operation. Based on the Haar wavelet, the DWT operation can be formulated as:
\begin{equation}
\begin{aligned}
    X^w_1 = L_hL_v(\hat{X}),  X^w_2 = L_hH_v(\hat{X}) \\
    X^w_3 = H_hL_v(\hat{X}),  X^w_4 = H_hH_v(\hat{X})    
\end{aligned}
,
\label{eq2}
\end{equation}
where the obtained frequency maps $X^w_1 \in \mathbb{R}^{B \times C \times h \times w}$, $X^w_2 \in \mathbb{R}^{B \times C \times h \times w}$ ,$X^w_3 \in \mathbb{R}^{B \times C \times h \times w}$, $X^w_4 \in \mathbb{R}^{B \times C \times h \times w}$ (here,$h = H/2$ and $w=W/2$) represent low-frequency approximation, horizontal, vertical, and diagonal details, respectively. Additionally, $L_h$ and $L_v$ are low-pass filter matrices and $H_h, H_v$ are high-pass filter matrices, which are defined by Haar wavelet basis:
\begin{equation}
    \begin{aligned}
        L=\frac{1}{\sqrt{2}} 
        \left[
            \begin{array}{cc}
                1 & 1 \\
                0 & 0
            \end{array}
        \right]
        , H=\frac{1}{\sqrt{2}}
        \left[
            \begin{array}{cc}
                1 & -1 \\
                0 & 0
            \end{array}
        \right]
    \end{aligned}
    . \label{eq3}
\end{equation}

\begin{figure}
    \centering
    \includegraphics[width=0.95\linewidth,height=4cm]{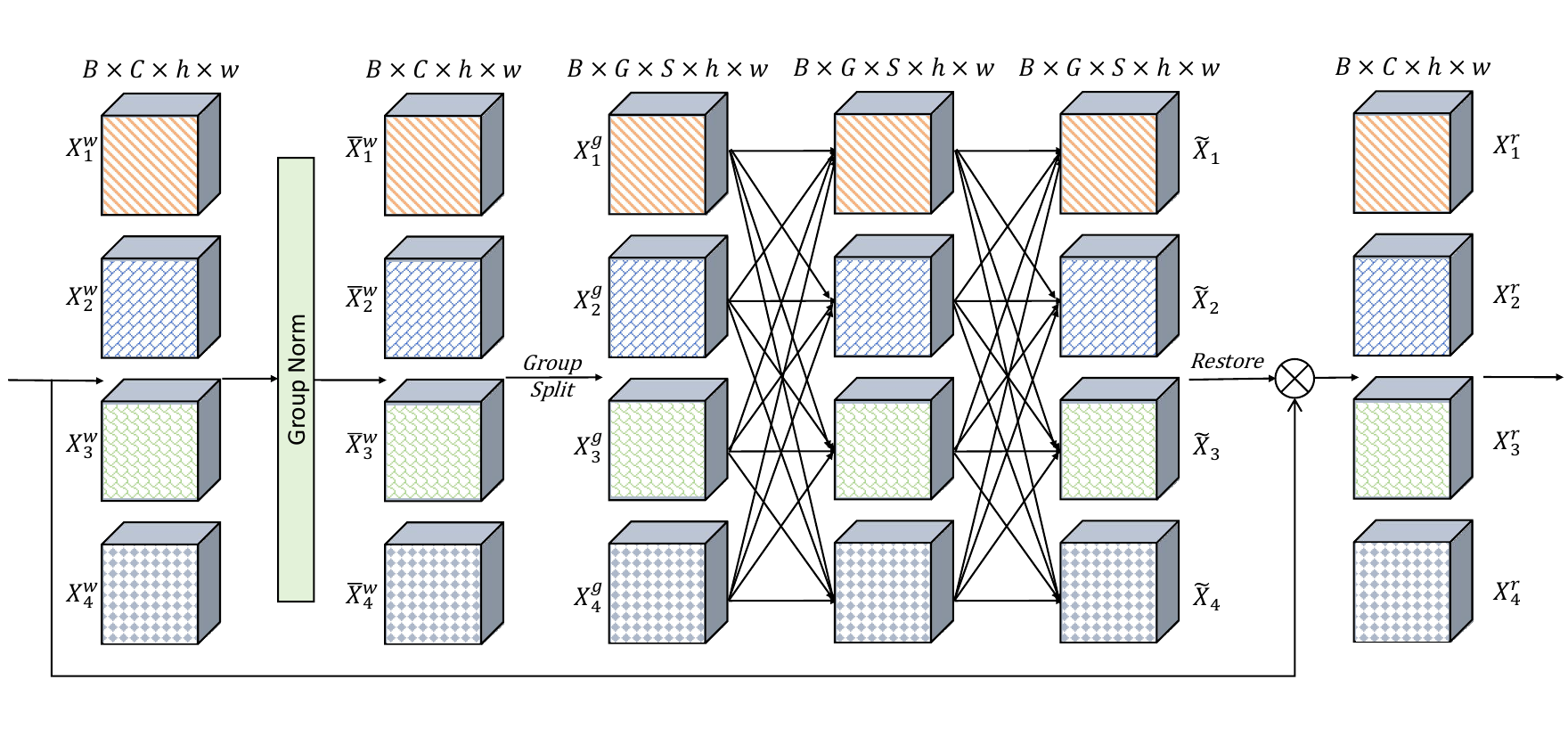}
    \caption{The simple implementation of adaptive group linear weighting operator.}
    \label{fig:AGLW}
\end{figure}

\paragraph{\textbf{Adaptive Group Linear Weighting}}
Considering that different frequency maps contain texture feature representations with varying levels of importance, this paper designs an Adaptive Group Linear Weighting (AGLW) operator to highlight informative texture feature representations and suppress redundant ones via long-range group interaction, as shown in Figure~\ref{fig:AGLW}. First, our AGLW applies group normalization (GroupNorm) operation to eliminate the inconsistencies of feature representations and reduce the redundancies among frequency maps. The formula of GroupNorm is written as follows:
\begin{equation}
\bar{X}^w_i = \text{GroupNorm}(X^w_i), \quad i \in \{1, 2, 3, 4\}.
\label{eq4}
\end{equation}

Next, to pursue a promising trade-off between computational efficiency and effectiveness, we spilt channels $C$ into $G$ independent groups, and each group has $S$ frequency maps (where $S=C/G$ is the size of each group). For example, frequency maps $X^w_1 \in \mathbb{R}^{B \times C \times h \times w}$ of low-frequency approximation are grouped into the corresponding grouped frequency maps  $X^g_1 \in \mathbb{R}^{B\times G \times S \times h\times w}$.

Then, the AGLW operator applies two successive linear transformations to model long-range group interactions for adjusting the relative significances of different frequency components:
\begin{equation}
\tilde{X}_i = W_i' \cdot \text{ReLU}(W_i \cdot X^g_i + b_i) + b_i', \quad i \in \{1, 2, 3, 4\},
\label{eq5}
\end{equation}
where $W_i, W_i' \in \mathbb{R}^{G\times S \times S}$, $b_i, b_i' \in \mathbb{R}^{G\times S}$, $\tilde{X}_i$ denote learnable weights and encoded token features.  

Finally, we restore the original channel dimension $C$ of frequency maps by multiplying augmented frequency maps with the original frequency maps through the element-wise product: 
\begin{equation}
X_i^{r} = \text{Restore}(\tilde{X}_i) \odot X^w_i, \quad i \in \{1, 2, 3, 4\},
\label{eq6}
\end{equation}
where $X_i^{r}$ denotes augmented frequency maps. In the ablation study, we will test the impacts of other weighting methods for the AGLW operator on the PD screening results.

\paragraph{\textbf{Inverse Wavelet Transform Reconstruction}}
The augmented frequency maps mainly comprise frequency components, which might degrade the PD screening performance. To address this issue, this paper develops an Inverse Wavelet Transform Reconstruction (IWTR) operator to reconstruct the refined feature maps with inverse wavelet transform operation:
\begin{equation}
    \begin{aligned}
        X^{out} &= \hat{X} + L^T_h[X_1^{r} + H_h^TX_2^{r}] + L_V^T[X_3^{r} + H_V^TX_4^{r}] \\
        Y &= Conv_{3\times3}(\text{Concat}[X, Conv_{1\times1}(X^{out})]),
        \label{eq7}
    \end{aligned}
\end{equation}
where $L^T_h, L^T_v, H^T_h, H^T_v$ are transpose matrices of the corresponding wavelet filters, $Y$ denotes the texture enhancement feature maps, $X^{out}$ denote the reconstructed feature maps, $Conv_{1\times1}$ represents $1 \times 1$ convolution for strengthening the representation diversity of $X^{out}$, $\text{Concat}$ denotes channel-wise concatenation operation, and $Conv_{3\times3}$ is the $3 \times 3$ convolution operation. The main function of $Conv_{3\times3}$ is to further enhance the informative texture feature representations, by fusing refined feature maps $\hat{X}$ and reconstructed feature maps $X^{out}$ dynamically via a residual skip connection.

\subsection{Balanced Confidence Loss}
Classical cross-entropy (CE) loss has widely been utilized to calculate loss values between ground truth and predictions. The formulation of CE Loss can be written as:
\begin{equation}
     L_{CE} = - \frac{1}{N} \sum_{i=1}^{N} \sum_{j=1}^{C} y_{ij}\log (p_{ij}), \label{eq9}
\end{equation}
where $N$ and $C$ denote the training sample number in a mini-batch and the class number, $y_{ij} \in \{0, 1\}$ and $p_{ij} \in \{p_{i1}, p_{i2},...,p_{ik},...,p_{iC}\}$ denote one-hot encoding of ground truth and 
predicting probability values generated by the classifier for different classes. However, CE treats each training sample equally for calculating the loss value in an averaging manner, bringing the issue of miscalibration/trustworthiness, especially for hard-classified example and rare classes. We explain this problem from the aspect of the gradient computing, and given the training samples $n_{k}$ with class $k$, its predicting probability value $p_{ij} = 
\frac{e^{z_{ij}}}{ {\textstyle \sum_{k=1}^{C} e^{z_{ik}}}}$. Thus, the gradient of class $k$ is computed as follows:
\begin{equation}
\frac{\partial L_{CE}}{\partial z_{mj}} =\begin{cases}
 &\frac{1}{N} \sum_{i=1}^{n_{j}} (p_{ij}-1) \text{ if } j=k \\
 &\frac{1}{N} \sum_{i=1}^{n_{j}}  p_{ik}  \text{ if } j\ne k
\end{cases},
 \label{eq10}
\end{equation}
where $n_{j}$ denote the number of training samples for class $j$. According to Equation~\ref{eq10}, we observe that head classes or easy-classified samples typically receive helpful gradients, while classes with few samples often receive suppressed gradients.

To tackle the above-mentioned issue, we develop a novel Balanced Confidence (BC) Loss, which aims to fully explore predicted probabilities and training label frequency distributions of different classes. The formula of BC is written as follows:
\begin{equation}
\begin{aligned}
    L_{BC} &= -\alpha L_{CS} + (1-\alpha)L_{CE} , \label{eq11}
\end{aligned}
\end{equation}
\begin{equation}
\begin{aligned}
    L_{CS} &=  -\frac{1}{N} \sum_{i=1}^{N} \sum_{j=1}^{C} y_{ij}\log(\hat{p}_{ij}) , \hat{p}_{ij} &= \frac{e^{z_{ij}}}{\sum_{i=1}^{N} \sum_{j \neq k}^{C} S_{kj} e^{z_{ik}} + e^{z_{ij}}}, \label{eq12}
\end{aligned}
\end{equation}
where $L_{CS}$ denotes the confidence boost loss,  $\alpha = 0.5$ is hyper-parameter for adjusting for the rebalance between sample-wise predicting probabilities and confidence scores, $\hat{p}_{ij}$ denote the modified predicting probability value for sample $i$ with class $j$, $S_{kj}$ is a balanced confidence factor to adjust predicted logits of other classes except for class $j$ by leveraging the potential of training label frequency prior and predicted probabilities of different classes. $S_{jk}$ is formulated as follows:
\begin{equation}
    S_{jk} = T_{jk} \cdot M_{jk},
   \label{eq12} 
\end{equation}
where $T_{jk}$ and $M_{jk}$ are two factors, which are determined by the prior frequency of the training label and the predicted probabilities of different classes accordingly:
\begin{equation}
    T_{kj} = 
    \begin{cases}
        1, & \text{if } p_{ik} \leq p_{ij} \\
        \left( \frac{p_{ik}}{p_{ij} }\right)^{\lambda}, & \text{if } p_{ik} > p_{ij}
    \end{cases},
     M_{kj} = 
    \begin{cases}
        1, & \text{if } n_{j} \leq n_{k}\\
        \left( \frac{n_j}{n_k}\right)^t, & \text{if } n_j > n_k
    \end{cases},
    \label{eq:13} 
\end{equation}

where $\lambda = 0.8$ and $t = 2$ are two hyper-parameters for adjusting the predicted logits of other classes except for the target class $j$, allowing classes with few examples or hard-classified samples to receive less negative gradients and obtain more positive gradients. Consequently, it also reduces the gap between the corresponding confidence scores and the accuracy of DNNs from the aspect of trustworthiness.  $n_j$ and $n_k$ represent the number of samples in classes $j$ and $k$, respectively.

\section{Experiments}

\subsection{Datasets}
In this paper, we employ two clinical OCT datasets of PD and a publicly available SD-OCT dataset of ocular diseases. Their introductions are described as follows:

\textbf{OCT-PD.}  It comprises 26 PD subjects and 49 healthy control (HC) subjects with a total number of 125 eyes. The number of OCT images (image resolution is $1561 \times 2048$) for each eye is 18; thus, the total number of OCT images in the OCT-PD dataset is 2250 (18 images/eye $\times$ 125): 1584 OCT images for HC (negative) and 666 OCT images for PD (positive). Here, we divide the OCT-PD dataset into five disjoint subsets at the subject level for five-fold cross-validation. In each fold, four subsets are used for training, and one subset is used for evaluation. Its collection is conducted according to the tenets of the Helsinki Declaration. 
This dataset can be available on reasonable request.

\textbf{WM-PD.} It is another OCT dataset of PD. The procedures for dataset collection are the same as the OCT-PD dataset. It comprises 59 PD subjects and 35 HC subjects, and only one eye was selected from each subject. WM-PD dataset has 1,692 OCT images with a resolution of $1378\times 2048$ (each eye contains 18 OCT images). The WM-PD dataset was split into three disjoint subsets based on the subject level: training (1,206 images: 450 images of HC and 756 images of PD), validation (306 images: 108 images of HC and 198 images of PD), and testing (180 images: 72 images of HC and 108 images of PD image).

\textbf{SD-OCT~\cite{srinivasan2014fully}.} It contains 45 subjects: 15 normal subjects, 15 subjects with age-related macular degeneration (AMD), and 15 subjects with diabetic macular edema (DME). We also split it into two disjoint subsets under the case level: training and testing. The training subset comprises 2,609 images: 917 with diabetic DME, 576 with dry AMD, and 1,116 normal. In this paper, 20\% of the training data are randomly selected at the case level for validation. The testing subset includes 621 images: 183 with diabetic DME, 147 with dry AMD, and 291 normal.

\subsection{Contrast Methods}
To investigate the effectiveness of our proposed AWFNet and BC, we adopt the following methods for comparisons:

\textbf{SOTA DNNs.} ResNet101~\cite{he2016deep}, ConvNeXt~\cite{Liu_2022_CVPR}, ViT~\cite{dosovitskiy2020image},  Swin-T~\cite{liu2021swin}, ResMLP~\cite{touvron2021resmlp}. Apart from these commonly-used DNNs, we utilized specific-designed frequency domain learning-based and medical image analysis-based DNNs for comparisons: WavePooling~\cite{williams2018wavelet}, Wavelet CNN (WCNN)~\cite{fujieda2018wavelet}, Deep adaptive wavelet network (DAWN)~\cite{Bastidas2019deep}, RIRNet~\cite{zhang2022attention}, FITNet~\cite{chen2023fit}, SRMNet~\cite{HUANG2024108368}, and RSGANet~\cite{shi2024retinal}.

\textbf{Competitive loss methods.}  FL~\cite{mukhoti2020calibrating},  D-Focal~\cite{tao2023dual}, and DCA~\cite{liang2020imporved}.


\subsection{Evaluation Metrics And Implementation Details}
\subsubsection{Evaluation Metrics}
In this paper, we comprehensively evaluate the general performance of methods from two perspectives: classification metrics and trustworthiness metrics.
 \textbf{Classification metrics:} Accuracy (ACC), precision (Pr), Sensitivity (Sen), F1 Score, Specificity (Spe), and balanced accuracy (b-ACC). Note that \textbf{ b-ACC} are a significant indictor to assess the performance of methods fairly on the imbalanced datasets by considering two clinical OCT datasets of PD are imbalanced. 
 
\textbf{Trustworthiness metrics:} Area under the curve (AUC), Expected Calibration Error (ECE), and Maximum Calibration Error (MCE)~\cite{tao2023dual}. AUC is vital for assessing the result trustworthiness of different methods clinically.

\subsubsection{Implementation Details}
We implement all DNNs and loss methods with PyTorch, Python, and OpenCV. All experiments are run on a machine with two NVIDIA GeForce RTX 2080 Ti GPUs. We employ Adam algorithm to optimize the learnable weights by using default settings. The initial learning rate was set to 0.0001. We use standard medical data augmentation strategies to preprocess training images, including random rotations and intensity normalization. Additionally, early stopping was implemented with a patience of 20 epochs and a minimum improvement of 1e-4. The input image size of three datasets are resized to $256 \times 256$.

\subsection{Result Analysis And Discussion}

\begin{table}
\caption{PD screening performance comparisons of AWF under three DNN stems.}
\centering
\resizebox{\columnwidth}{!}{
\label{tab1}
\begin{tabular}{ccccc}
\hline
Method& ACC & F1 & b-ACC   \\
\hline
ResNet18~\cite{he2016deep} & $0.7430 \pm 0.0671$ & $0.2586 \pm 0.1491$ & $0.5769 \pm 0.0778$  \\
+AWF& $\mathbf{0.8444 \pm 0.0115}$ & $\mathbf{0.3751 \pm 0.0908}$ & $\mathbf{0.6219 \pm 0.0453}$ \\ \hline
ResNet101~\cite{he2016deep} & $0.7792 \pm 0.0326$ & $0.2493 \pm 0.1167$ & $0.5648 \pm 0.0368$ \\
+AWF& $\mathbf{0.8145\pm0.0365}$ & $\mathbf{0.3653\pm0.0657}$ & $\mathbf{0.6133\pm0.0405}$ \\ \hline
WCNN~\cite{fujieda2018wavelet} & $0.7816 \pm 0.0240$ & $\mathbf{0.2955 \pm 0.0519}$ & $0.5795 \pm 0.0245$  \\
+AWF& $\mathbf{0.8043\pm0.0209}$ & $0.2852\pm0.0883$ &$\mathbf{0.5811\pm0.0374}$\\
\hline
AIM~\cite{yang2023aim}& $0.7899\pm0.0344$ & $0.2451\pm0.1597$&$0.5757±0.0495$\\
+AWF& $\bm{0.8333\pm0.0059}$ & $0.2499\pm0.1405$& $\bm{0.5768\pm0.0554}$\\
\hline
\end{tabular}
}
\end{table}

\begin{table*}
\caption{PD screening performance comparisons of our AWFNet and state-of-the-art deep neural networks on the PD-OCT dataset.}
\centering
\resizebox{1.96\columnwidth}{!}{
    \begin{tabular}{ccccccc}
    \hline
    Network &  ACC & F1 & Sen & Pr & Spe & b-ACC  \\
  \hline
    ResNet18~\cite{he2016deep} & $0.7430 \pm 0.0671$ & $0.2586 \pm 0.1491$ & $0.3222 \pm 0.2333$ & $0.3369 \pm 0.1464$ & $0.8316 \pm 0.1156$ & $0.5769 \pm 0.0778$  \\
    ConvNeXt~\cite{Liu_2022_CVPR} &$0.8174 \pm 0.0184$ & $0.0991 \pm 0.0813$ & $0.0639 \pm 0.0524$ & $0.2671 \pm 0.2641$ & $\bm{0.9760 \pm 0.0290}$ & $0.5200 \pm 0.0185$  \\
    ViT~\cite{dosovitskiy2020image} &$0.8261 \pm 0.0000$ & $0.0000 \pm 0.0000$ & $0.0000 \pm 0.0000$ & $0.0000 \pm 0.0000$ & $1.0000 \pm 0.0000$ & $0.5000 \pm 0.0000$  \\ 
    Swin-T~\cite{liu2021swin}&$0.8068\pm0.0364$ & $0.1006\pm0.1314$&$0.0806\pm0.1011$&$0.1417\pm0.2012$&$0.9596\pm0.0557$&$0.5201\pm0.0376$\\
    ResMLP~\cite{touvron2021resmlp}&$0.8019\pm0.0483$ & $0.0000\pm0.0000$&$0.0000\pm0.0000$&$0.0000\pm0.0000$&$0.9708\pm0.0585$&$0.4854\pm0.0292$\\
    \hline \hline
    WavePooling~\cite{williams2018wavelet}&$0.7986\pm0.0357$ & $0.0536\pm0.0559$&$0.0417\pm0.0512$&$0.1067\pm0.0879$&$0.9579\pm0.0540$&$0.4998\pm0.0022$\\
    WCNN~\cite{fujieda2018wavelet} &$0.7816\pm 0.0240$&$0.2955\pm 0.0519$&$0.2694\pm0.0717$&$0.3523\pm0.0593$&$0.8895\pm0.0389$&$0.5795\pm0.0245$\\
    DAWN~~\cite{Bastidas2019deep} &$0.8261 \pm 0.0000$ & $0.0000 \pm 0.0000$ & $0.0000 \pm 0.0000$ & $0.0000 \pm 0.0000$ & $1.0000 \pm 0.0000$ & $0.5000 \pm 0.0000$ \\ \hline \hline
    RIRNet~\cite{zhang2022attention}&$0.8043\pm0.0162$ & $0.3276\pm0.1729$&$\bm{0.3389\pm0.2213}$&$0.5337\pm0.2342$&$0.9023\pm0.0635$&$0.6206\pm0.0796$\\
    FITNet~\cite{chen2023fit}&$0.7575\pm0.0556$ & $0.2687\pm0.1386$&$0.3111\pm0.1883$&$0.2589\pm0.1345$&$0.8515\pm0.1052$&$0.5813\pm0.0448$\\
    SRMNet~\cite{HUANG2024108368}&$0.8256\pm0.0169$ & $0.2875\pm0.1813$&$0.2389\pm0.1688$&$0.3890\pm0.2131$&$0.9491\pm0.0318$&$0.5940\pm0.0719$\\
  
    RSGANet~\cite{shi2024retinal}&$0.8338\pm 0.0136$&$0.2899\pm 0.1219$&$0.2139\pm 0.1269$&$0.6181\pm 0.1695$&$0.9643\pm 0.0287$&$0.5891\pm 0.0522$\\ \hline \hline
    \textbf{AWFNet} &$\bm{0.8444 \pm 0.0115}$ & $\bm{0.3751 \pm 0.0908}$ & $0.2806 \pm 0.1045$ & $\bm{0.6300 \pm 0.0694}$ & $0.9632 \pm 0.0189$ & \bm{$0.6219 \pm 0.0453$} \\
    \hline
    \end{tabular}
    }
    \label{tab2}
\end{table*}

\begin{table*}
    \centering
    \caption{Performance comparisons of our BC and advanced loss methods in terms of PD screening performance and trustworthiness on the PD-OCT dataset.}
     \resizebox{\textwidth}{!}{
    \begin{tabular}{c|cccccc|ccc}
        \hline
        \multirow{2}{*}{Method} &\multicolumn{6}{c|}{Screening metrics} & \multicolumn{3}{c}{Trustworthiness metrics}\\
       & ACC & F1 & Sen & Pr & Spe & b-ACC & AUC & ECE & MCE \\
        \hline
          CE & $0.8444\pm0.0115$ & $0.3751\pm0.0908$&$0.2806\pm0.1045$&$0.6300\pm0.0694$&$0.9632\pm0.0189$&$0.6219\pm0.0453$&$0.7228\pm0.1310$&$0.1373\pm0.0658$&$0.2568\pm0.1304$\\ 
          FL~\cite{mukhoti2020calibrating} &$0.8145\pm0.0105$ & $0.3226\pm0.1139$&$0.2750\pm0.1265$&$0.4479\pm0.0507$&$0.9281\pm0.0287$&$0.6015\pm0.0508$&$0.6962\pm0.0966$&$0.1356\pm0.0624$&$0.2393\pm0.0706$\\  
          D-Focal~\cite{tao2023dual}&$0.8303\pm0.0513$ & $0.0472\pm0.0405$&$0.0444\pm0.0416$&$0.0653\pm0.0689$&$0.9329\pm0.0628$&$0.4886\pm0.0159$&$0.5252\pm0.0420$&$0.1375\pm0.0434$&$0.3281\pm0.0742$\\
          DCA~\cite{liang2020imporved}&$0.8038\pm0.0252$ & $0.0907\pm0.0897$&$0.0963\pm0.0962$&$0.0868\pm0.0848$&$0.8961\pm0.0337$&$0.4962\pm0.0409$&$0.4558\pm0.0678$&$0.1393\pm0.0345$&$0.2463\pm0.0684$\\
          CS &$0.8222\pm0.0109$ & $0.2413\pm0.2181$&$0.2333\pm0.2542$&$0.3807\pm0.2743$&$0.9462\pm0.0551$&$0.5898\pm0.1006$&$0.6799\pm0.1410$&$0.1521\pm0.0893$&$0.2823\pm0.0467$\\  
          BC &$\bm{0.8531\pm0.0151}$ & $\bm{0.4242\pm0.1125}$&$\bm{0.3306\pm0.1271}$&$\bm{0.6864\pm0.0916}$&$\bm{0.9632\pm0.0266}$&$\bm{0.6469\pm0.0540}$&$\bm{0.7796\pm0.0653}$&\bm{$0.1330\pm0.0616$}&$\bm{0.2311\pm0.0442}$\\  
        \hline
    \end{tabular}
    }
    \label{tab3}
\end{table*}

\subsubsection{PD Screening Performance Comparisons Under Different DNN Stems}
Table~\ref{tab1} presents the PD screening results of the AWF based on the PD-OCT dataset across Five different network stems: ResNet18, ResNet101,  WCNN, and AIM~\cite{yang2023aim}. Our AWF consistently improves PD screening performance through comparisons to three original DNN stems. Specifically, WCNN is a frequency domain learning-based DNN, and  AIM is a vision foundation model.
For example, compared with ResNet18, AWF obtains \textbf{10.14\%}, \textbf{11.65\%}, and 4.5\% gains of accuracy, F1, and b-ACC, respectively. WCNN achieves higher F1 than it with AWF but performs worse results of accuracy and b-ACC. It is worth noting that AWF under ResNet18 obtains better PD screening results than it with other DNN stems; thus, we take ResNet18 as the DNN stem for our AWF module in the following experiments.
The results verify the superiority of AWF as a practical texture feature amplifier in mining the informative texture representations in high-level feature representations for boosting the PD screening performance.

\subsubsection{PD Screening Performance Comparisons With SOTA DNNs} 
Table~\ref{tab2} provides the PD screening performance of our AWFNet and existing advanced DNNs on the PD-OCT dataset. It can be observed that ConvNeXt obtains the highest specificity value, but it gets quite poor results of F1, sensitivity, precision, and b-ACC, indicating it has a bias in screening HC and PD, which is not applicable to automated PD screening. Compared with Swin-T, AWFNet obtains absolute over gains of 3.76\%, \textbf{27.45\%}, \textbf{48.83\%}, \textbf{10.18\%} gains of accuracy, F1, precision, and b-ACC. Remarkably, AWFNet significantly outperforms WCNN by \textbf{6.28\%} of accuracy, \textbf{7.96\%} of F1, \textbf{27.77\%} precision, and 4.24\% b-ACC. 
Compared to specific-designed RSGANet and SRMNet of OCT-based PD screening, AWFNet obtains gains of \textbf{8.52\%} in F1, \textbf{4.17\%} in sensitivity, and 2.79\% b-ACC. The results manifest the helpful texture feature representation enhancement in high-level feature maps of DNNs improves PD screening results from the frequency domain learning aspect, aligning with our motivation.

\subsubsection{Comparisons To SOTA Loss Methods} 
Table~\ref{tab3} offers the PD screening results and trustworthiness of our BC and comparable loss methods based on our AWFNet. Compared to FL and D-Focal, BC achieves absolute over 2.28\%, 10.16\%, 5.56\%, 3.03\%, 4.54\%, and 8.34\% improvements in ACC, F1, sensitivity, specificity, b-ACC, and AUC, while slightly decreasing ECE and MCE. BC significantly outperforms DCA by 4.93\% in ACC, 33.35\% in F1, \textbf{23.43\%} in sensitivity, \textbf{59.96\%} in precision, 6.71\% in specificity, 14.87 in b-ACC, and \textbf{32.02\%} in AUC, as well as a reduction of 1.52\% in MCE. From the aspect of b-ACC, our BC predicts PD and HC more fairly than other loss methods, which is essential for clinical diagnosis.
In general, the results show the superiority of our BC in efficiently exploring the predicted probabilities of different classes and their frequency priori.

\subsubsection{PD Screening Performance Analysis From Age-related Aspect}

Here, we divide subjects of PD and HC into two age groups based on the PD-OCT dataset: Group$<$ 60 and Group$\ge$60. The reason for adopting the age of 60 as the threshold is that the incidence of PD increases rapidly over 60. Table~\ref{tab5-2} shows that AWFNet achieves higher accuracy, F1, precision, specificity, and b-ACC than other competitive DNNs under the Group$<$ 60. For example, AWFNet outperforms SMRNet by absolute over 3.97\%, 4.39\%, 21.91\%, and 4.82\% of accuracy, F1, precision, and specificity accordingly. However, three trustworthiness metrics of AWFNet are not as promising as we expected. AWFNet with BC keeps a better trade-off between PD screening and trustworthiness. Regarding the result analysis of Group$\ge$ 60, AWFNet also generally performs better than other comparable DNNs. It with BC further boosts PD screening and trustworthiness performance. Overall, the results verify that our AWFNet with BC recognizes PD and HC credibly and accurately without ageism by treating AWF as the efficient texture feature amplifier and the predicted probabilities of different classes, meeting clinical diagnosis requirements.

\begin{table*}
\caption{Performance comparisons of representative DNNs and our proposed methods in terms of PD screening performance and trustworthiness from the the age-related aspect.}
 \label{tab5-2}
\centering
\resizebox{1.99\columnwidth}{!}{
\begin{tabular}{c|cccccc|ccc}
    \hline
    \multirow{2}{*}{Method} &\multicolumn{6}{c|}{Screening metrics} & \multicolumn{3}{c}{Trustworthiness metrics}\\
     & ACC & F1 & Sen & Pr & Spe & b-ACC & AUC & ECE & MCE \\
    \hline
     \multicolumn{10}{c}{ Group$<$ 60 } \\ \hline
    ResNet18~\cite{he2016deep} & $0.7619\pm0.1202$ & $0.1644\pm0.2342$ & $0.2111\pm0.2731$ & $0.1402\pm0.2114$ & $0.8537\pm0.1433$ & $0.5324\pm0.1391$ & $0.6418\pm0.1899$ & $\bm{0.1312\pm0.0424}$ & $0.2946\pm0.0400$ \\
    Swin-T~\cite{liu2021swin}&$0.8190\pm0.0723$ & $0.1055\pm0.1553$ & $0.1000\pm0.1333$ & $0.1207\pm0.1938$ & $0.9389\pm0.0921$ & $0.5194\pm0.0613$ & $0.5577\pm0.1179$ & $0.1396\pm0.0722$ & $\bm{0.2894\pm0.1185}$\\
    WCNN~\cite{fujieda2018wavelet}&$0.7857\pm0.0317$ & $0.1200\pm0.1118$ & $0.1111\pm0.0994$ & $0.1359\pm0.1381$ & $0.8981\pm0.0388$ & $0.5046\pm0.0472$ & $0.5772\pm0.1515$ & $0.1352\pm0.0197$ & $0.3930\pm0.1662$\\ 
    SRMNet~\cite{HUANG2024108368}&$0.8524\pm0.0129$ & $0.2669\pm0.2402$ & $0.2778\pm0.2811$ & $0.3142\pm0.2612$ & $0.9481\pm0.0519$ & $0.6130\pm0.1159$ & $0.7767\pm0.1511$ & $0.1431\pm0.0390$ & $0.4177\pm0.2182$\\
    \textbf{AWFNet} & $0.8921\pm0.0556$ & $0.3108\pm0.3703$ & $0.2667\pm0.3758$ & \bm{$0.5333\pm0.4522$} & \bm{$0.9963\pm0.0045$} & $0.6315\pm0.1890$ & $0.7147\pm0.2879$ & $0.1455\pm0.0525$ & $0.4050\pm0.1280$ \\
    \textbf{AWFNet+BC} &$\bm{0.8984\pm0.0622}$ & $\bm{0.3543\pm0.4373}$ & $\bm{0.3222\pm0.4043}$ & $0.4000\pm0.4899$ & $0.9944\pm0.0111$ & $\bm{0.6583\pm0.2044}$ & $\bm{0.8848\pm0.0985}$ & $0.1357\pm0.0952$ & $0.2913\pm0.1247$\\
    \hline 
   \multicolumn{10}{c}{ Group$\ge$60 } \\ \hline
    ResNet18 & $0.7347\pm0.0536$ & $0.2920\pm0.1436$ & $\bm{0.3593\pm0.2362}$ & $0.3578\pm0.1317$ & $0.8214\pm0.1122$ & $0.5903\pm0.0704$ & $0.7119\pm0.0840$ & $0.1433\pm0.0535$ & $0.3542\pm0.1590$ \\
    Swin-T~\cite{liu2021swin}&$0.8014\pm0.0208$ & $0.1000\pm0.1255$ & $0.0741\pm0.0915$ & $0.1593\pm0.2106$ & $\bm{0.9692\pm0.0392}$ & $0.5217\pm0.0322$ & $0.6212\pm0.0233$ & $0.1126\pm0.0404$ & $0.2593\pm0.1415$\\
    WCNN~\cite{fujieda2018wavelet} &$0.7799\pm0.0352$ & $0.3499\pm0.0603$ & $0.3222\pm0.0897$ & $0.4268\pm0.1124$ & $0.8855\pm0.0562$ & $0.6038\pm0.0309$ & $0.7283\pm0.0674$ & $0.1185\pm0.0512$ & $0.2535\pm0.1417$\\ 
    SRMNet~\cite{HUANG2024108368}&$0.8139\pm0.0218$ & $0.2867\pm0.1686$ & $0.2259\pm0.1393$ & $0.4069\pm0.2262$ & $0.9496\pm0.0318$ & $0.5877\pm0.0606$ & $0.7090\pm0.0822$ & $0.1191\pm0.0516$ & $0.3922\pm0.2129$\\
    \textbf{AWFNet} & $0.8236\pm0.0218$ & $0.3770\pm0.0504$ & $0.2852\pm0.0477$ & $0.5878\pm0.1190$ & $0.9479\pm0.0290$ & $0.6165\pm0.0227$ & $0.7259\pm0.0896$ & $0.1135\pm0.0258$ & $0.3261\pm0.1516$\\
    \textbf{AWFNet+BC} &$\bm{0.8333\pm0.0394}$ & $\bm{0.4267\pm0.1169}$ & $0.3333\pm0.1034$ & $\bm{0.6461\pm0.1855}$ & $0.9487\pm0.0408$ & $\bm{0.6410\pm0.0567}$ & $\bm{0.7412\pm0.0510}$ &\bm{ $0.1071\pm0.0483$} & \bm{$0.2524\pm0.0688$}\\

    \hline 
    \end{tabular}}
\end{table*}

\subsubsection{Generalization And Robustness Validation}

\begin{table*}
\centering
\caption{Result comparisons of our methods and SOTA DNNs on SD-OCT and WM-PD.}
\begin{tabular}{l|l|ccccc|ccc}
\hline
\multirow{2}{*}{Dataset} & \multirow{2}{*}{Method} &\multicolumn{5}{c|}{Classification metrics} & \multicolumn{3}{c}{Trustworthiness metrics}\\
  &   & ACC & Pr & Sen & F1 & b-ACC & AUC & ECE & MCE \\ 
\hline
\multirow{13}{*}{SD-OCT}
& ResNet18~\cite{he2016deep} & $0.9372$ & $0.9415$ & $0.9363$ & $0.9349$ & $0.9363$ & $0.9965$ & $0.0266$ & $0.2993$ \\
& ConvNeXt~\cite{Liu_2022_CVPR} & $0.5556$ & $0.6210$ & $0.4989$ & $0.5068$ & $0.4989$ & $0.7072$ & $0.1002$ & $0.3625$ \\
& ViT~\cite{dosovitskiy2020image} & $0.8052$ & $0.7935$ & $0.7509$ & $0.7562$ & $0.7509$ & $0.9232$ & $0.1354$ & $0.3548$ \\

& Swin-T~\cite{liu2021swin}&$0.7472$ & $0.7585$ & $0.7237$ & $0.7342$ & $0.7237$ & $0.8892$ & $0.1714$ & $0.4538$\\

& ResMLP~\cite{touvron2021resmlp}&$0.5298$ & $0.6065$ & $0.4590$ & $0.4631$ & $0.4590$ & $0.6544$ & $0.0784$ & $0.3793$\\
& WavePooling~\cite{williams2018wavelet}&$0.4944$ & $0.5195$ & $0.4414$ & $0.4467$ & $0.4414$ & $0.5810$ & $0.0639$ & $0.4569$\\
& WCNN~\cite{fujieda2018wavelet} & $0.9098$ & $0.9212$ & $0.8943$ & $0.8976$ & $0.8943$ & $0.9969$ & $0.0353$ & $0.6491$ \\
& DAWN~\cite{Bastidas2019deep} & $0.7568$ & $0.7658$ & $0.7204$ & $0.7261$ & $0.7204$ & $0.8543$ & $0.0843$ & $\bm{0.2260}$ \\
& RIRNet~\cite{zhang2022attention}&$0.8068$ & $0.8669$ & $0.7581$ & $0.7768$ & $0.7581$ & $0.9626$ & $0.1463$ & $0.5711$\\
& FITNet~\cite{chen2023fit}&$0.8953$ & $0.8880$ & $0.9152$ & $0.8978$ & $0.9152$ & $0.9863$ & $0.0758$ & $0.5098$\\
& SRMNet~\cite{HUANG2024108368}&$0.8277$ & $0.8633$ & $0.8269$ & $0.8281$ & $0.8269$ & $0.9726$ & $0.0967$ & $0.3957$\\ \cline{2-10}
& AWFNet & $0.9662$ & $0.9697$ & $0.9678$ & $0.9688$ & $0.9678$ & $\bm{0.9976}$ &$0.0222$ & $0.7587$\\ 
& AWFNet+BC&$\bm{0.9742}$ & $\bm{0.9757}$ & $\bm{0.9769}$ & $\bm{0.9763}$ & $\bm{0.9769}$ & $0.9969$ & $\bm{0.0130}$ & $0.5179$\\
\hline \hline
\multirow{13}{*}{WM-PD}
& ResNet18~\cite{he2016deep} & $0.7722$ & $0.7860$ & $0.7361$ & $0.7448$ & $0.7361$ & $0.8185$ & $0.0975$ & $0.5036$\\
& ConvNeXt~\cite{Liu_2022_CVPR} & $0.6000$ & $0.5506$ & $0.5023$ & $0.3876$ & $0.5023$ & $0.4394$ & $0.1789$ & $0.4767$ \\
& ViT~\cite{dosovitskiy2020image} & $0.5667$ & $0.5000$ & $0.5000$ & $0.4665$ & $0.5000$ & $0.6421$ & $0.2034$ & $0.4293$ \\
& Swin-T~\cite{liu2021swin}&$0.5889$ & $0.5966$ & $0.9722$ & $0.7394$ & $0.4931$ & $0.7503$ & $0.1725$ & $0.4345$\\
& ResMLP~\cite{touvron2021resmlp}&$0.5833$ & $0.6279$ & $0.7500$ & $0.6835$ & $0.5417$ & $0.6037$ & $0.3376$ & $0.3938$\\
& WavePooling~\cite{williams2018wavelet}&$0.5778$ & $0.5909$ & $0.9630$ & $0.7324$ & $0.4815$ & $0.6780$ & $0.2559$ & $0.4835$\\
& WCNN~\cite{fujieda2018wavelet}& $0.7778$ & $0.8214$ & $0.7315$ & $0.7408$ & $0.7315$ & $0.8808$ & $0.1502$ & $0.3858$ \\
& DAWN~\cite{Bastidas2019deep} & $0.6056$ & $0.8017$ & $0.5069$ & $0.3900$ & $0.5069$ & $0.5900$ & $0.0930$ & $0.0930$ \\
& RIRNet~\cite{zhang2022attention}&$0.5889$ & $0.6000$ & $0.9444$ & $0.7338$ & $0.5000$ & $0.6466$ & $0.3894$ & $0.4095$\\
& FITNet~\cite{chen2023fit}&$0.6778$ & $0.6712$ & $0.9074$ & $0.7717$ & $0.6204$ & $0.7325$ & $0.3072$ & $0.5077$\\
& SRMNet~\cite{HUANG2024108368}&$0.6333$ & $0.6265$ & $0.9630$ & $0.7591$ & $0.5509$ & $0.6952$ & $0.1844$ & $0.4015$\\ \cline{2-10}
& AWFNet & $0.8167$ & $0.7778$ & $\bm{0.9722}$ & $0.8642$ & $0.7778$ & $0.8897$ & $0.1614$ & $0.5077$ \\
& AWFNet+BC&$\bm{0.8333}$ & $\bm{0.8362}$ & $0.8981$ & $\bm{0.8661}$ & $\bm{0.8171}$ & $\bm{0.8969}$ & $\bm{0.0571}$ & $\bm{0.2369}$\\
\hline
\end{tabular}
\label{tab: different dataset}
\end{table*}

\textbf{Result analysis on SD-OCT.} As listed at the top of Table~\ref{tab: different dataset}, AWFNet generally achieves better retinal disease classification performance than SOTA DNNs on the SD-OCT dataset. Remarkably,  AWFNet significantly outperforms Swin-T by \textbf{21.90\%} in accuracy, \textbf{21.12\%} in precision, \textbf{24.41\%} in sensitivity, \textbf{23.46\%} in F1, \textbf{24.41\%} in b-ACC, and \textbf{10.84\%} in AUC, while reducing ECE by \textbf{14.92\%}. Compared to FITNet and SRMNet, AWFNet obtains improvements of \textbf{13.85\%} in accuracy, \textbf{10.64\%} in precision, \textbf{14.09\%} in sensitivity, \textbf{14.07\%} in F1, \textbf{14.09\%} in b-ACC, and a 2.50\% increase in AUC, as well as reducing ECE by \textbf{5.36\%}. AWFNet with BC further boosts retinal disease classification and trustworthiness performance.

\textbf{Result analysis on WM-PD.} As presented at the bottom of Table~\ref{tab: different dataset} on the WM-PD dataset, AWFNet also gets better PD screening results than other SOTA DNNs. Notably, AWFNet outperforms ConvNeXt by absolute over \textbf{21.00\%} in accuracy, \textbf{26.16\%} in precision, \textbf{46.77\%} in sensitivity, \textbf{48.66\%} in F1, \textbf{25.75\%} in b-ACC, and \textbf{44.59\%} in AUC, while reducing ECE by \textbf{17.20\%}. AWFNet also outperforms ResMLP by \textbf{23.34\%} in accuracy, \textbf{22.72\%} in precision, \textbf{46.99\%} in sensitivity, \textbf{47.66\%} in F1, \textbf{27.55\%} in b-ACC, and \textbf{45.03\%} in AUC, while reducing ECE by \textbf{17.62\%}. Compared with WCNN, DAWN, and RIRNet, AWFNet obtains gains of \textbf{22.78\%} in accuracy, \textbf{17.78\%} in precision, \textbf{46.53\%} in sensitivity, \textbf{47.42\%} in F1 score, \textbf{27.78\%} in b-ACC, and \textbf{29.97\%} in AUC, as well as reducing ECE by \textbf{22.80\%}. AWFNet with BC further improves  PD screening performance and trustworthiness, demonstrating its generalization and robustness. Overall, the results on these two datasets verify the effectiveness and generalization abiltiy of AWFNet in extracting useful texture feature representations in high-level feature maps and BC in leveraging the potential of predicted probabilities of different classes and their frequencies.

\begin{table*}
    \centering
    \caption{PD screening result comparisons of  AWF with/without AWF-Mixer and Channel Mixer}
    \resizebox{\textwidth}{!}{
    \begin{tabular}{cccccccc}
    \hline
      &  ACC & F1 & Sen  & Spe & b-ACC  \\
    \hline
    ResNet18 & $0.7430 \pm 0.0671$ & $0.2586 \pm 0.1491$ & $0.3222 \pm 0.2333$ &  $0.8316 \pm 0.1156$  & $0.5769 \pm 0.0778$ \\
    Channel Mixer & $0.7739\pm0.0425$ & $0.3013\pm0.1597$&$\bm{0.3417\pm0.2084}$&$0.8649\pm0.0910$&$0.6033\pm0.0629$\\
   Channel Mixer+AWF-Mixer &$\bm{0.8444 \pm 0.0115}$ & $\bm{0.3751 \pm 0.0908}$ & $0.2806 \pm 0.1045$  & $\bm{0.9632 \pm 0.0189}$  & $\bm{0.6219 \pm 0.0453}$  \\
  \hline
    \end{tabular}
    }
    \label{tab:dwt}
\end{table*}

\begin{table*}
    \centering
    \caption{Effects of AWF number in AWFNet.}
    \resizebox{\textwidth}{!}{
    \begin{tabular}{c|ccccccc}
 \hline
    Number of AWFs & ACC & F1 & Sen  & Spe & b-ACC\\
    \hline
    0 &$0.7430\pm0.0671$ & $0.2586\pm0.1491$&$0.3222\pm0.2333$&$0.8316\pm0.1156$&$0.5769\pm0.0778$\\
    1 & $0.8106\pm0.0355$ & $0.3156\pm0.0778$&$0.2778\pm0.1584$&$0.9228\pm0.0749$&$0.6003\pm0.0435$\\
    2 & $0.8145\pm0.0174$ & $0.2199\pm0.1936$&$0.2111\pm0.2096$&$0.9415\pm0.0641$&$0.5763\pm0.0731$ \\ 
    3 &$\bm{0.8444\pm0.0115}$ & $\bm{0.3751\pm0.0908}$&$0.2806\pm0.1045$&$\bm{0.9632\pm0.0189}$&$\bm{0.6219\pm0.0453}$ \\
    4&$0.7686\pm0.0594$ & $0.2394\pm0.1720$&$\bm{0.2889\pm0.2441}$&$0.8696\pm0.1196$&$0.5792\pm0.0670$\\
    5&$0.8053\pm0.0321$ & $0.2948\pm0.1516$&$0.2750\pm0.1618$&$0.9170\pm0.0671$&$0.5960\pm0.0525$\\
   \hline
    \end{tabular}
    }
    \label{tab:10}
\end{table*}


\begin{figure*}
    \centering
    \begin{minipage}{0.23\linewidth}
        \centering
        \includegraphics[width=0.95\linewidth,height=3cm]{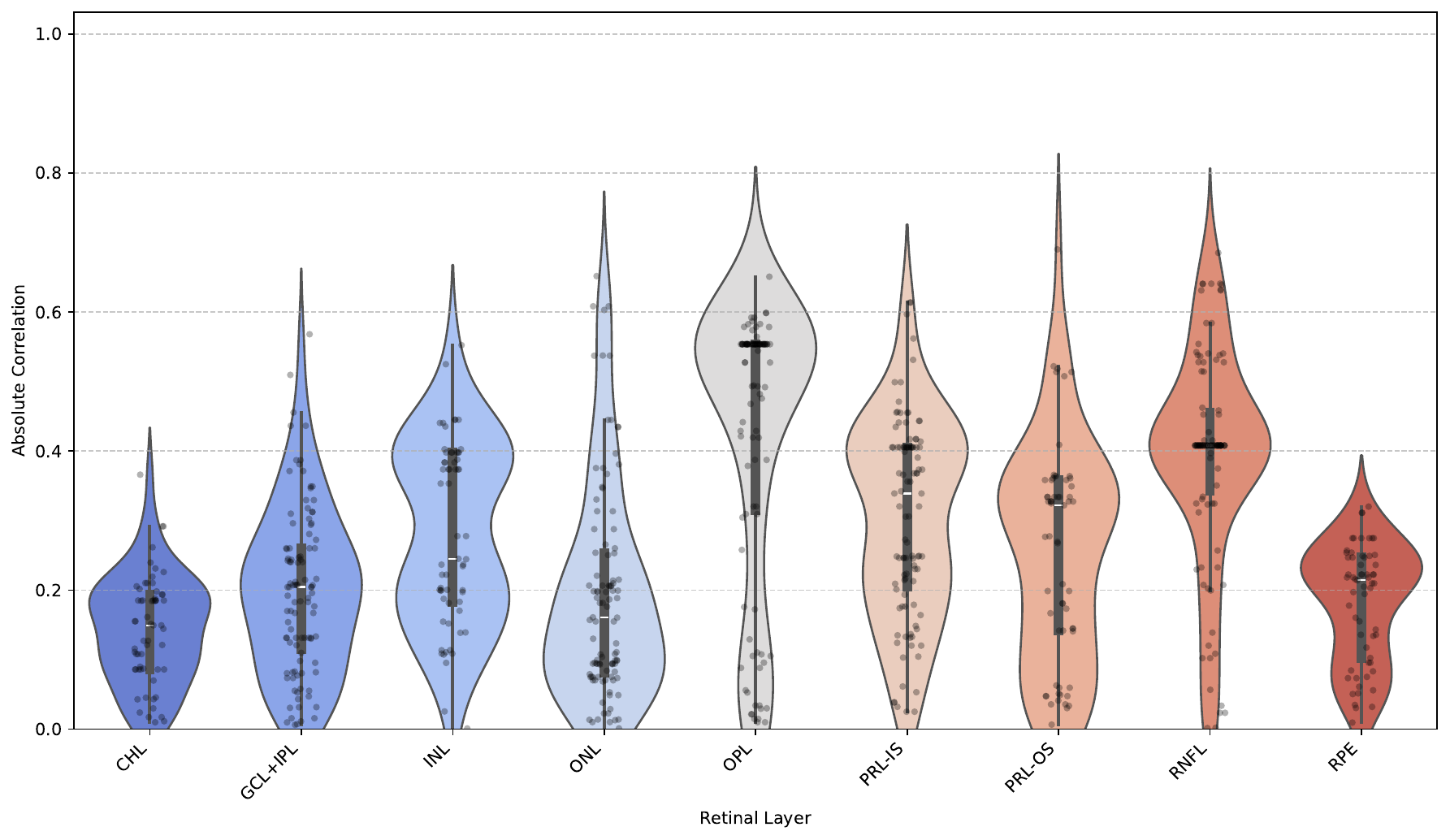}
    \end{minipage}
    \begin{minipage}{0.23\linewidth}
        \centering
        \includegraphics[width=0.95\linewidth,height=3cm]{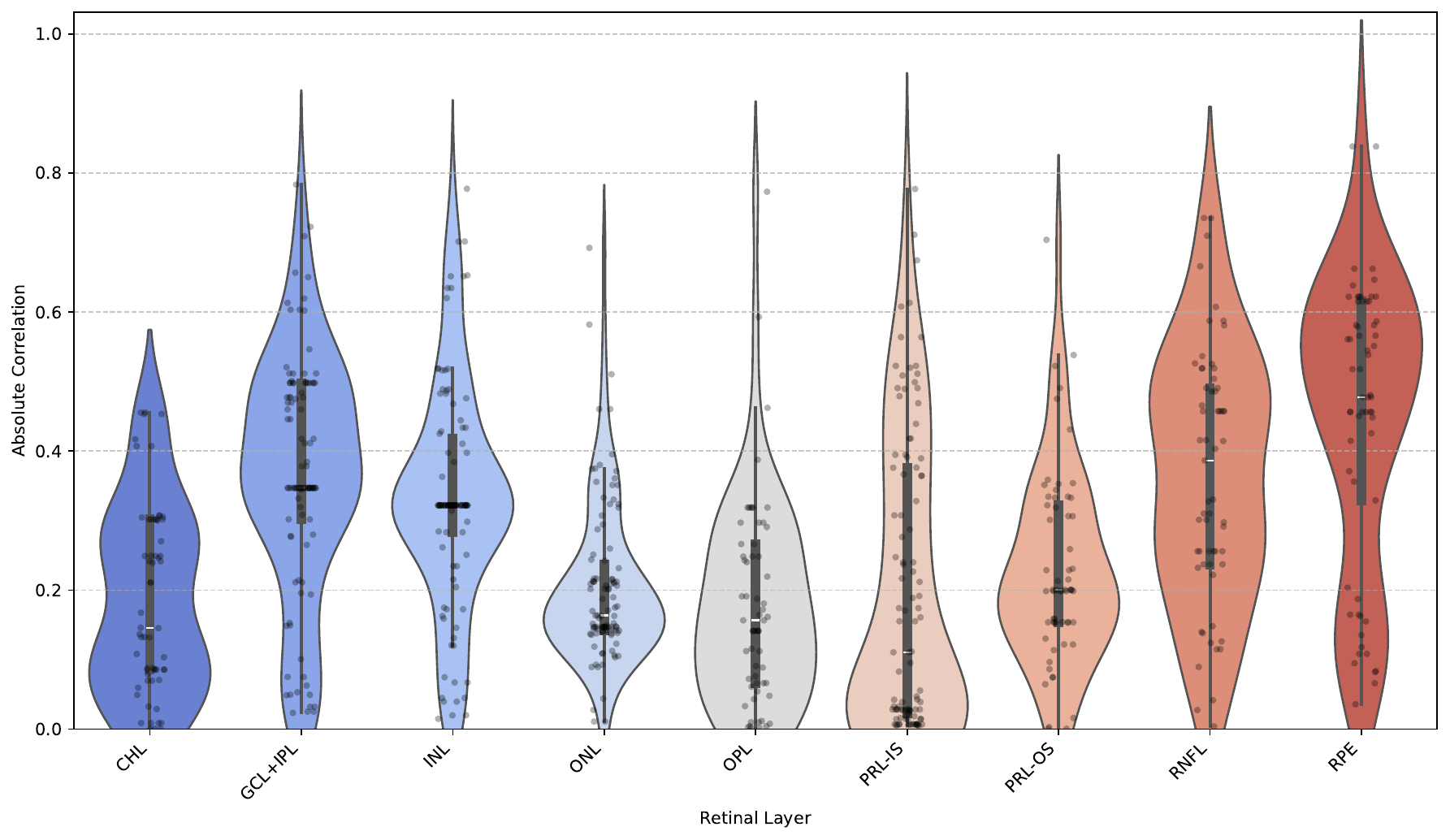}
    \end{minipage}
    \begin{minipage}{0.23\linewidth}
        \centering
        \includegraphics[width=0.95\linewidth,height=3cm]{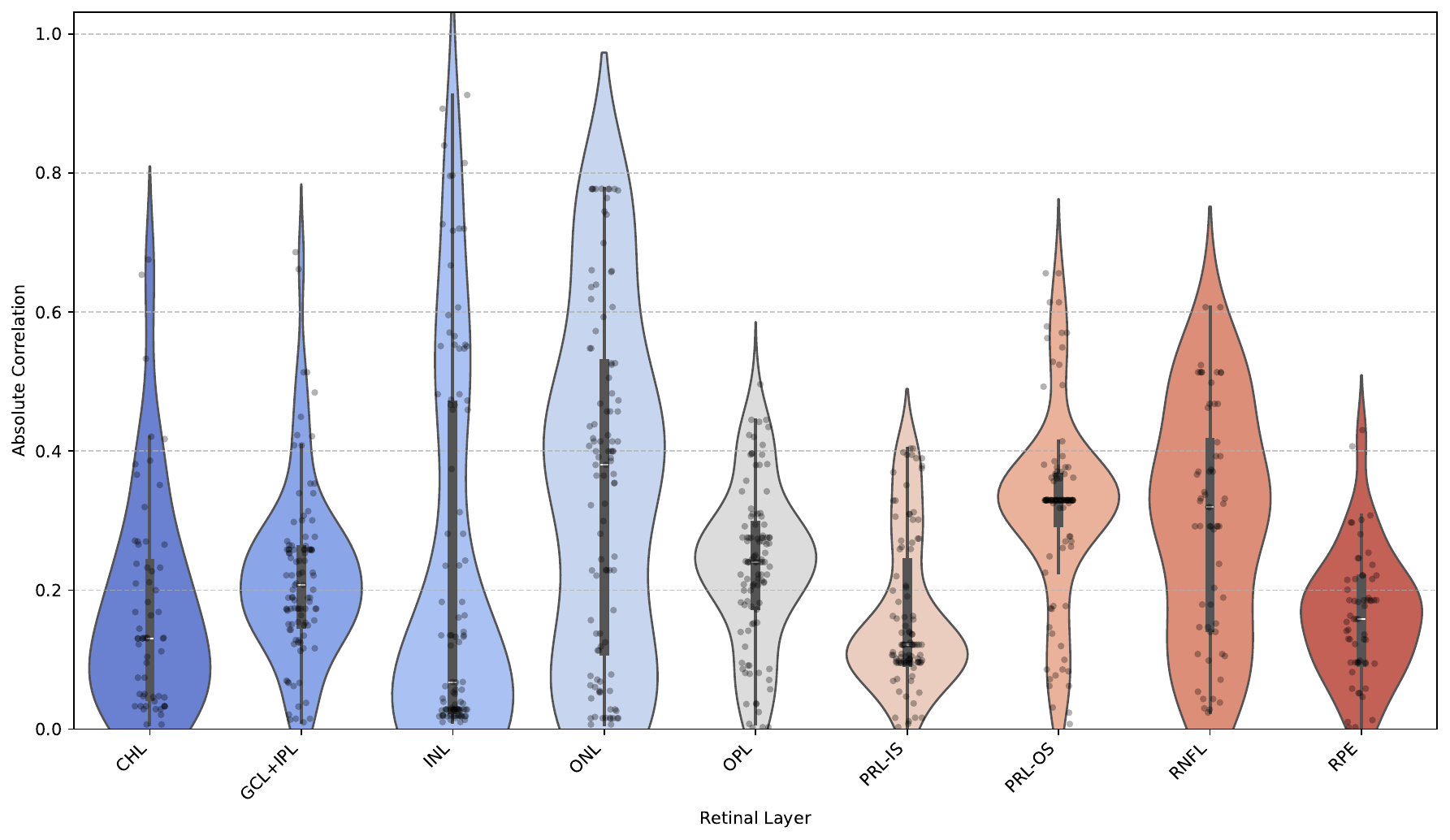}
    \end{minipage}
    \begin{minipage}{0.23\linewidth}
        \centering
    \includegraphics[width=0.95\linewidth,height=3cm]{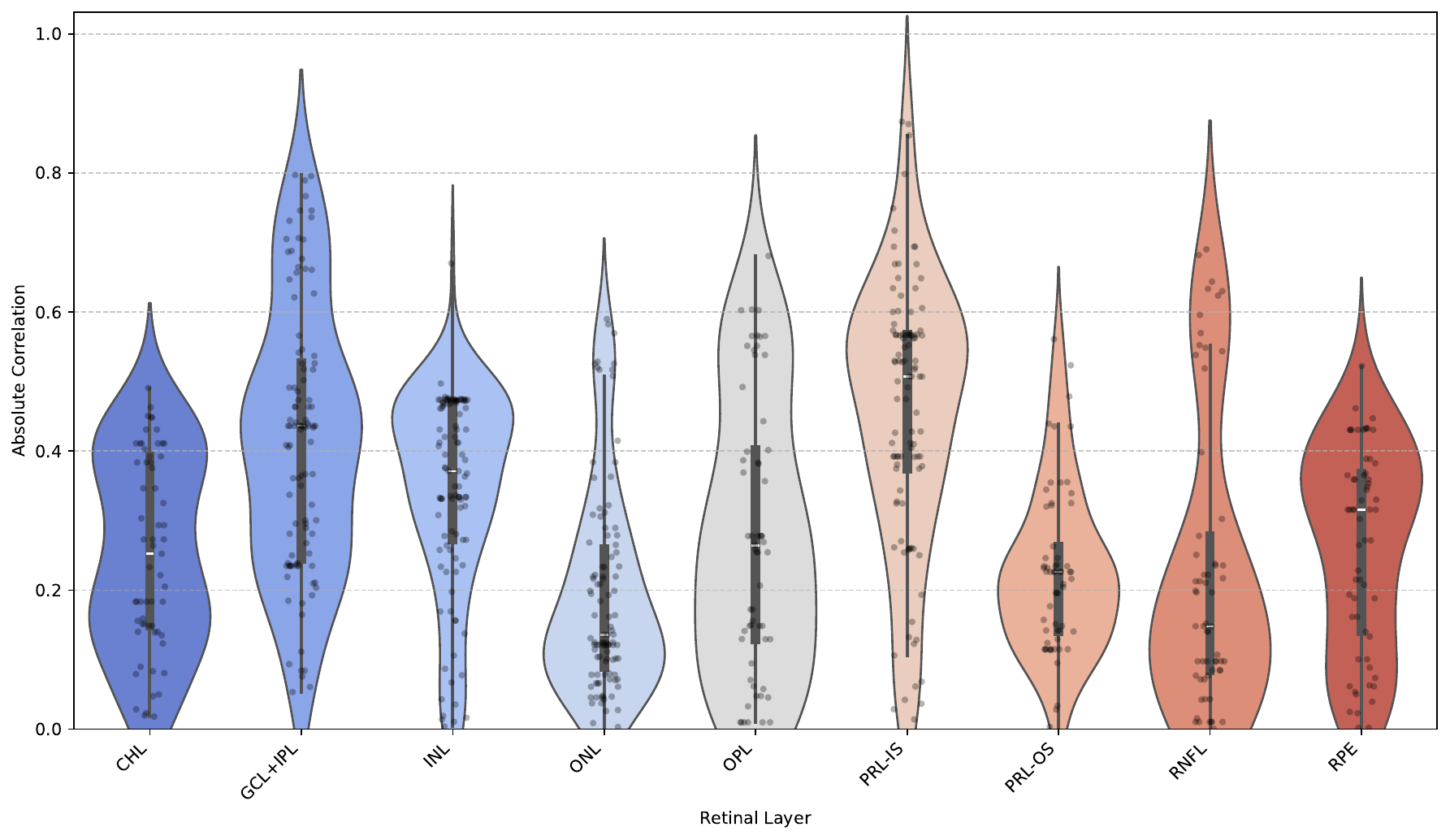}
    \end{minipage}
    \caption{The visualizations of the correlation coefficients among texture features extracted from different retinal layers based on OCT images and the frequency components extracted from high-level feature maps, which are obtained from different DNNs ((ResNet18, ViT, SRMNet, and AWFNet, left-to-right order) accordingly. Here, the vertical axis denotes correlation coefficient distribution, and the horizontal axis denotes nine different retinal layers.}
    \label{fig8}
\end{figure*}

\subsection{Ablation Study}

\subsubsection{Effects of AWF-Mixer And Channel Mixer In AWF}
Table~\ref{tab:dwt} provides the result comparisons of our AWF with/without AWF-Mixer and channel mixer under ResNet18. AWF with the AWF-Mixer significantly outperforms AWF without the AWF-Mixer and ResNet18 by absolute over \textbf{7.05}\% of accuracy, \textbf{7.38\%} of F1, \textbf{9.83}\% of specificity, respectively. Additionally, AWF without the AWF-Mixer also performs better than ResNet18, showing the channel mixer also enhances the PD screening performance via feature representation refinement. The results manifest the significant roles of AWF-Mixer and channel mixer in the PD screening results.

\subsubsection{Effects Of The Number Of AWFs}
Table~\ref{tab:10} presents the result comparisons of our AWFNet with different number of AWFs. We observe that the number of AWF modules plugged at the end of the DNN stem that have significant impacts on PD screening performance. It is challenging to set the proper number of AWFs. Our AWFNet with three AWFs performs better than other settings.

\subsection{Visualization Analysis}
Fig.~\ref{fig8} presents correlation coefficient distribution visualizations among texture features and frequency components via violin plots. Here, texture features are extracted from original OCT images, and frequency components are extracted from high-level feature maps of four DNNs (ResNet18, ViT, SRMNet, and AWFNet) via Fast Fourier Transforms (FFT). The vertical axis indicates correlation coefficient distributions among texture features extracted from the specific retinal layer and various frequency components, and the horizontal axis indicates different retinal layers. We conduct a log transformation for original correlation coefficient values to highlight the differences visually. It can be observed that compared with correlation coefficient value distributions of ResNet18,  the corresponding values of our AWFNet significantly increase, indicating our method is capable of enhancing the extraction of texture feature representations by treating the AWF as the practical texture feature amplifier. Additionally, our AWFNet generally highlights the texture feature representations of different retinal layers better than ViT and SRMNet, supporting our hypothesis.

\section{Conclusion and future work}
Inspired by the great potential of texture features as biomarkers for clinical PD diagnosis, we present the Adaptive Wavelet Filtering Network (AWFNet) to improve the PD screening performance, by effectively mining the texture feature representations from high-level feature maps with the well-designed AWF as the efficient texture feature amplifier. Additionally, to further improve PD screening results and the trustworthiness of AWFNet, we develop a balanced confidence loss by leveraging predicted probabilities of different classes and training label frequency priori. The extensive experiments manifest the superiority of our methods over SOTA DNNs and loss methods in terms of PD screening metrics and trustworthiness metrics. Further, age-related analysis and visualizations manifest the effectiveness of our proposed methods. In the future, we plan to modify our AWFNet and BC to boost PD screening performance and trustworthiness and deploy it on real medical devices to test its applicability and robustness.

\bibliographystyle{IEEEtran}

\bibliography{mybibliography}

\newpage

 




\vfill

\end{document}